\documentclass[prl,twocolumn,twoside,preprintnumbers,superscriptaddress,nofootinbib]{revtex4}

\usepackage{amsmath,slashed}
\usepackage{graphicx,graphics,color}
\usepackage{dcolumn}
\usepackage[hyperfootnotes=false]{hyperref}
\usepackage{xspace}
\usepackage{enumerate}
\usepackage{varwidth}

\usepackage{physics}
\usepackage{amssymb}
\usepackage{mathtools}
\usepackage{dsfont}
\usepackage{multirow}

\hypersetup{
    colorlinks=true,
    linkcolor=blue,
    citecolor=blue
}

\newcommand{\mpi}{M_\pi}

\newcommand{\beq}{\begin{equation}}
\newcommand{\eeq}{\end{equation}}
\newcommand{\diff}{\text{d}}
\newcommand{\eps}{\epsilon}

\newcommand{\Order}{\mathcal{O}}

\newcommand{\GeV}{\,\text{GeV}}

\renewcommand{\Im}{\text{Im}\,}

\newcommand{\etapp}{\eta^{(\prime)}}
\newcommand{\sm}{s_\text{m}}

\allowdisplaybreaks[1]

\begin{document}

\title{Precision evaluation of the $\boldsymbol{\eta}$- and $\boldsymbol{\eta'}$-pole contributions to hadronic\\[1mm] light-by-light scattering in the anomalous magnetic moment of the muon}

\author{Simon Holz}
\affiliation{Albert Einstein Center for Fundamental Physics, Institute for Theoretical Physics, University of Bern, Sidlerstrasse 5, 3012 Bern, Switzerland}
\author{Martin Hoferichter}
\affiliation{Albert Einstein Center for Fundamental Physics, Institute for Theoretical Physics, University of Bern, Sidlerstrasse 5, 3012 Bern, Switzerland}
\author{Bai-Long Hoid}
\affiliation{Albert Einstein Center for Fundamental Physics, Institute for Theoretical Physics, University of Bern, Sidlerstrasse 5, 3012 Bern, Switzerland}
\affiliation{Institut f\"ur Kernphysik and PRISMA$^+$  Cluster of Excellence, Johannes Gutenberg Universit\"at,  55099 Mainz, Germany}  
\author{Bastian Kubis}
\affiliation{Helmholtz-Institut f\"ur Strahlen- und Kernphysik (Theorie) and
Bethe Center for Theoretical Physics, Universit\"at Bonn, 53115 Bonn, Germany}

\begin{abstract} 
 Next to the $\pi^0$ pole, $\eta$ and $\eta'$ intermediate states give rise to the leading singularities of the hadronic light-by-light tensor, resulting in sizable contributions to the anomalous magnetic moment of the muon $a_\mu$. The strength of the poles is determined by the respective transition form factors (TFFs) to two (virtual) photons. We present a calculation of these TFFs that implements a number of low- and high-energy constraints, including the $\etapp\to\gamma\gamma$ decay widths, $\etapp\to\pi^+\pi^-\gamma$ spectra, chiral symmetry for the $\etapp\to2(\pi^+\pi^-)$ amplitudes, vector-meson couplings, and asymptotic limits. Crucially, we investigate the role of the leading left-hand singularity generated by the exchange of the $a_2$ tensor meson, yielding, for the first time, an estimate of the associated factorization-breaking corrections. Our final results, $a_\mu^{\eta\text{-pole}}=14.7(9)\times 10^{-11}$ and $a_\mu^{\eta'\text{-pole}}=13.5(7)\times 10^{-11}$, conclude a dedicated effort to evaluate the pseudoscalar-pole contributions to hadronic light-by-light scattering using dispersion relations, amounting to a combined $a_\mu^{\text{PS-poles}}=91.2^{+2.9}_{-2.4}\times 10^{-11}$.
\end{abstract}

\maketitle

\emph{Introduction}---The anomalous magnetic moment of the muon $a_\mu$ has been measured at the Fermilab experiment to a precision of $0.20$ ppm~\cite{Muong-2:2023cdq,Muong-2:2024hpx}, already dominating the world average
\beq
a_\mu^\text{exp}=116\,592\,059(22)\times 10^{-11},
\eeq
and further improvements with Runs 4+5+6 will likely allow the collaboration to surpass the design goal of $0.14$ ppm~\cite{Muong-2:2015xgu}.
While currently no Standard-Model prediction of comparable precision is available, efforts are ongoing~\cite{Colangelo:2022jxc} to update the previous white-paper consensus~\cite{Aoyama:2020ynm,Aoyama:2012wk,Aoyama:2019ryr,Czarnecki:2002nt,Gnendiger:2013pva,Davier:2017zfy,Keshavarzi:2018mgv,Colangelo:2018mtw,Hoferichter:2019gzf,Davier:2019can,Keshavarzi:2019abf,Hoid:2020xjs,Kurz:2014wya,Melnikov:2003xd,Colangelo:2014dfa,Colangelo:2014pva,Colangelo:2015ama,Masjuan:2017tvw,Colangelo:2017qdm,Colangelo:2017fiz,Hoferichter:2018dmo,Hoferichter:2018kwz,Gerardin:2019vio,Bijnens:2019ghy,Colangelo:2019lpu,Colangelo:2019uex,Blum:2019ugy,Colangelo:2014qya}, and eventually achieve a commensurate level. In this context, the most urgent challenge concerns the resolution of tensions in the evaluation of hadronic vacuum polarization, both within data-driven approaches~\cite{Davier:2017zfy,Keshavarzi:2018mgv,Colangelo:2018mtw,Hoferichter:2019gzf,Davier:2019can,Keshavarzi:2019abf,Hoid:2020xjs,Crivellin:2020zul,Keshavarzi:2020bfy,Malaescu:2020zuc,Colangelo:2020lcg,Stamen:2022uqh,Colangelo:2022vok,Colangelo:2022prz,Hoferichter:2023sli,Hoferichter:2023bjm,Stoffer:2023gba,Davier:2023fpl,Leplumey:2025kvv} (most notably in the context of the CMD-3 measurement of $e^+e^-\to\pi^+\pi^-$~\cite{CMD-3:2023alj,CMD-3:2023rfe} and potentially related to radiative corrections~\cite{Campanario:2019mjh,Ignatov:2022iou,Colangelo:2022lzg,Monnard:2021pvm,Abbiendi:2022liz,BaBar:2023xiy,Aliberti:2024fpq}) and with lattice QCD~\cite{Borsanyi:2020mff,Ce:2022kxy,ExtendedTwistedMass:2022jpw,FermilabLatticeHPQCD:2023jof,RBC:2023pvn,Boccaletti:2024guq,Blum:2024drk,Djukanovic:2024cmq,Bazavov:2024eou}. At the same time, however, also the precision of hadronic light-by-light (HLbL) scattering needs to be improved by at least a factor $2$, because otherwise the current uncertainty, $a_\mu^\text{HLbL} = 92(19) \times 10^{-11}$~\cite{Aoyama:2020ynm,Melnikov:2003xd,Masjuan:2017tvw,Colangelo:2017qdm,Colangelo:2017fiz,Hoferichter:2018dmo,Hoferichter:2018kwz,Gerardin:2019vio,Bijnens:2019ghy,Colangelo:2019lpu,Colangelo:2019uex,Pauk:2014rta,Danilkin:2016hnh,Jegerlehner:2017gek,Knecht:2018sci,Eichmann:2019bqf,Roig:2019reh}, would limit the sensitivity. Efforts in this direction are ongoing both in lattice QCD~\cite{Blum:2019ugy,Chao:2021tvp,Chao:2022xzg,Blum:2023vlm,Fodor:2024jyn} and with data-driven techniques, the latter focusing on the role of subleading effects related to hadronic states at intermediate energies~\cite{Hoferichter:2020lap,Zanke:2021wiq,Danilkin:2021icn,Stamen:2022uqh,Ludtke:2023hvz,Hoferichter:2023tgp,Hoferichter:2024fsj,Ludtke:2024ase,Deineka:2024mzt,Hoferichter:2025yih}, higher-order short-distance constraints~\cite{Bijnens:2020xnl,Bijnens:2021jqo,Bijnens:2022itw,Bijnens:2024jgh}, and their implementation~\cite{Leutgeb:2019gbz,Cappiello:2019hwh,Knecht:2020xyr,Masjuan:2020jsf,Ludtke:2020moa,Colangelo:2021nkr,Leutgeb:2021mpu,Leutgeb:2022lqw,Colangelo:2024xfh,Leutgeb:2024rfs}.

At the same time, the evaluation of the leading contribution to HLbL scattering, from the light pseudoscalar poles $P=\pi^0, \eta, \eta'$, cannot be considered fully satisfactory yet either. For the $\pi^0$, a comprehensive dispersive analysis of the crucial transition form factor (TFF) for $\pi^0\to\gamma^*\gamma^*$ has been performed~\cite{Hoferichter:2018dmo,Hoferichter:2018kwz}, incorporating all available low- and high-energy constraints, and the result agrees with determinations using Canterbury approximants~\cite{Masjuan:2017tvw} and lattice QCD~\cite{Gerardin:2019vio}.\footnote{A potential deficit between more recent lattice-QCD calculations~\cite{Christ:2022rho,Gerardin:2023naa,ExtendedTwistedMass:2023hin,Lin:2024khg} and the experimental normalization from $\pi^0\to\gamma\gamma$~\cite{PrimEx-II:2020jwd} might require further investigation.} For $\eta$ and $\eta'$, however, the evaluation from Ref.~\cite{Aoyama:2020ynm} is entirely based on Canterbury approximants, since only recently first lattice-QCD calculations have become available~\cite{ExtendedTwistedMass:2022ofm,Gerardin:2023naa} and no full dispersive evaluation existed to date. In this Letter, we present such an analysis, as vital input for a complete dispersive analysis of HLbL scattering~\cite{Hoferichter:2024vbu,Hoferichter:2024bae,Hoferichter:2025fea}.

To this end, we build upon previous work~\cite{Stollenwerk:2011zz,Hanhart:2013vba,Kubis:2015sga,Holz:2015tcg,Holz:2022hwz,Holz:2022smu} reconstructing $\etapp$ TFFs from dispersion relations, augmented by several crucial advances. First, we combine insights from chiral perturbation theory (ChPT) to construct $\etapp\to2(\pi^+\pi^-)$ amplitudes that, together with the exchange of an $a_2$ tensor meson as the dominant left-hand cut, allow for a good description of the $\etapp\to\pi^+\pi^-\gamma$ decay spectra, and thus can serve as seeds for the dispersive reconstruction of the isovector component of the TFFs. Second, we perform a  dispersive unitarization of these amplitudes, accounting for the complicated cut structure due to $a_2$ exchange, which allows us, for the first time, to assess the size of factorization-breaking contributions. Third, at each step in the construction we implement constraints on the asymptotic behavior, both for possible subtractions of the dispersive integrals and regarding the high-energy behavior of the resulting TFFs. All parts of the calculation are supplemented with detailed uncertainty estimates, and the results are validated against space-like data for $e^+e^-\to e^+e^-\etapp$ as well as slope parameters measured, e.g., in $\etapp\to e^+e^-\gamma$ decays. With the isoscalar component determined by measured vector-meson decays and normalizations from $\etapp\to\gamma\gamma$, we thus obtain TFF representations that comply with constraints from analyticity and unitarity, various data for $\etapp$ decays and production, chiral symmetry, and perturbative QCD. This translates to robust predictions for the $\etapp$-pole contributions to HLbL scattering at an unprecedented level of precision.
 
\begin{figure}[tb]
     \centering
     \includegraphics[width=\linewidth]{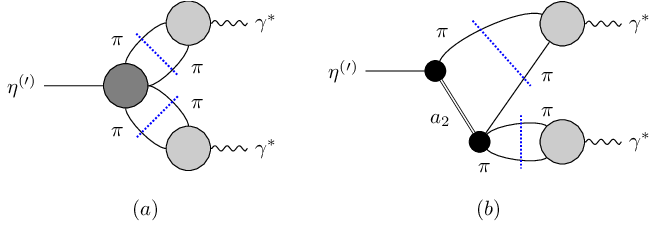}
     \caption{Diagrammatic representation of the isovector $\etapp$ TFFs $(a)$ with its left-hand-cut contribution $(b)$. Dotted lines indicate unitarity cuts.}
     \label{fig:isovector_diags}
 \end{figure}

\emph{Outline of the calculation}---The pseudoscalar TFFs describe the decay $P(q_1+q_2)\to\gamma^*(q_1,\mu)\gamma^*(q_2,\nu)$, formally defined by the matrix elements
\begin{align}
&i \int  \diff^4x\, e^{iq_1 \cdot x} \langle 0 | T\{ j_{\mu}(x)j_{\nu}(0) \}  | P(q_1+q_2) \rangle \notag\\
&=  \epsilon_{\mu\nu\rho\sigma}q_1^{\rho}q_2^{\sigma} F_{P\gamma^*\gamma^*}(q_1^2,q_2^2),
\end{align}
where the normalization $F_{P\gamma^*\gamma^*}(0,0)\equiv F_{P\gamma\gamma}$ is determined by the $P\to\gamma\gamma$ decay rate. 
The TFFs appear as residues of the pseudoscalar poles in the HLbL tensor, whose contributions to $a_\mu$ can be evaluated with standard techniques~\cite{Knecht:2001qg,Knecht:2001qf,Jegerlehner:2009ry}. In our calculation, we 
decompose the $\etapp$ TFFs according to
 \beq
     F_{\etapp \gamma^*\gamma^*} = F_{\etapp}^{(I=1)} + F_{\etapp}^{(I=0)} + F_{\etapp}^{\text{eff}} + F_{\etapp}^{\text{asym}}, \label{Eq:tff_compl}
 \eeq
 in analogy to the $\pi^0$ case~\cite{Hoferichter:2018dmo,Hoferichter:2018kwz,Hoferichter:2021lct}. The main difference to similar analyses for the $\pi^0$~\cite{Schneider:2012ez,Hoferichter:2012pm,Hoferichter:2014vra} concerns the isospin structure: since $I=0$ for $\etapp$, either both photons have isovector or isoscalar quantum numbers, leading to two independent terms in Eq.~\eqref{Eq:tff_compl}. The first term, the isovector dispersive piece, reproduces all the lowest-lying singularities in this channel, while also taking into account factorization-breaking effects, see Fig.~\ref{fig:isovector_diags} for the two main topologies. The required  
 $\etapp \to 2(\pi^+\pi^-)$ amplitude~\cite{Guo:2011ir} is constructed including constraints from ChPT, with the resulting $\eta'$ partial width in this channel~\cite{BESIII:2014bgm,BESIII:2023ceu} as another cross-check. Factorization breaking is first introduced by the exchange of the tensor resonance $a_2(1320)$, since $P$-wave contributions are suppressed: the quantum numbers, $J^{PC}=1^{-+}$, only permit an exotic $\pi_1(1600)$ resonance~\cite{ParticleDataGroup:2024cfk,COMPASS:2014vkj,CrystalBarrel:2019zqh}, at higher energy than the $a_2(1320)$. Moreover, the $P$-wave discontinuity in $\pi\eta$ scattering only appears at $\Order(p^8)$ in ChPT, further suppressing the low-energy tail, so that the dominant left-hand-cut structure is generated by the $D$-wave.   
 
 The parameters of the $\etapp\to2(\pi^+\pi^-)$ amplitudes are fixed by data fits to the pion spectra in the decays $\etapp \to \pi^+ \pi^- \gamma$~\cite{KLOE:2012rfx,BESIII:2017kyd}, with cross-checks by a narrow-resonance approximation against specialized dispersive representations of the same amplitudes~\cite{Stollenwerk:2011zz,Kubis:2015sga}.
 Furthermore, the couplings are cross-checked against the outcome of phenomenological Lagrangian models~\cite{Bando:1987br,Giacosa:2005bw,Ecker:2007us} for combinations of the partial decay widths $a_2 \to \etapp \pi$, $a_2 \to 3\pi$, $a_2\to\pi\gamma$, and $\rho \to \pi^+ \pi^-$~\cite{ParticleDataGroup:2024cfk}. The unitarization of these amplitudes, indicated by the unitarity diagrams in Fig.~\ref{fig:isovector_diags}, proceeds via dispersion relation as discussed below.

 The second term in Eq.~\eqref{Eq:tff_compl} parameterizes the small contribution of narrow resonances in the isoscalar channel, with their couplings fixed phenomenologically from the partial decay widths for $\omega\to \eta \gamma$, $\eta' \to \omega \gamma$, $\phi\to\etapp \gamma$, and $\omega,\, \phi \to e^+ e^-$~\cite{ParticleDataGroup:2024cfk} and their signs from $U(3)$ symmetry~\cite{Hanhart:2013vba,Gan:2020aco}. The third term provides a parameterization of higher intermediate states, to ensure exact agreement with the $\etapp\to\gamma\gamma$ decay width and high-energy, singly-virtual data on $e^+e^-\to e^+e^-\etapp$. The last term implements asymptotic constraints from perturbative QCD~\cite{Lepage:1979zb,Lepage:1980fj,Brodsky:1981rp}, including corrections due to the finite $\etapp$ masses~\cite{Zanke:2021wiq,Holz:2024diw}. Decay constants and mixing angles for the $\eta$--$\eta'$ system from lattice QCD~\cite{Ottnad:2017bjt,Bali:2021qem} provide additional cross-checks on the asymptotic coefficients. In this Letter, we report the main results of our calculation, while a detailed  description is given in Ref.~\cite{Holz:2024diw}.

 \emph{$\eta$ and $\eta'$ transition form factors}---The dispersive reconstruction of the dominant low-energy piece proceeds via its $2\pi$ singularities, see Fig.~\ref{fig:isovector_diags}$(a)$. The required $\etapp\to2(\pi^+\pi^-)$ amplitude is calculated based on a hidden-local-symmetry ansatz~\cite{Guo:2011ir}, which ensures that the amplitude behaves as $\Order(p^6)$ in the chiral expansion, and supplemented by an $a_2$ contribution, see Fig.~\ref{fig:isovector_diags}$(b)$. For the dispersive unitarization, in the first step the $\etapp \to \pi^+ \pi^- \gamma^*$ partial-wave amplitude is found as
 \begin{align}
 \mathcal{F}_{\etapp \pi\pi\gamma}(t,k^2) = \frac{1}{96 \pi^2} &\int_{4 M_\pi^2}^{\Lambda^2} \diff x\, \frac{x \sigma_\pi^3(x) \big[F_\pi^V(x)\big]^*}{x-k^2-i\eps} \label{Eq:etapipig}\\
 &\times \Big[f_1(t,x)\Omega(x) + f_1(x,t) \Omega(t) \Big], \notag
 \end{align}
 where $k^2$ is the photon virtuality, $t$ the $2\pi$ invariant mass squared, $\sigma_\pi(s)=\sqrt{1-4 \mpi^2/s}$, $F_\pi^V$ the pion vector form factor, $\Omega$ the Omn\`es function~\cite{Omnes:1958hv}, and $f_1(s,t)$ the partial-wave amplitude of the decay into four pions. To ensure the correct asymptotic behavior, this dispersion relation is kept unsubtracted. This implies, however, that the sum rule for $F_{\etapp \gamma \gamma}$ is violated at a level around $10\, \%$, a mismatch to be restored using $F_{\etapp}^{\text{eff}}$ below. For the numerical evaluation of Eq.~\eqref{Eq:etapipig},
 we use $F_\pi^V(s)=P(s)\Omega(s)$, where the parameters of the (linear) polynomial $P$ are fit to the data of Ref.~\cite{Belle:2008xpe}.
 
\begin{figure}[tb]
     \centering
     \includegraphics[width=\linewidth]{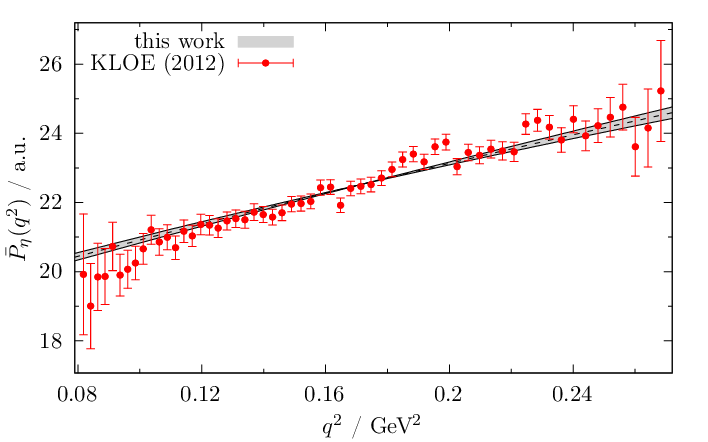}
     \includegraphics[width=\linewidth]{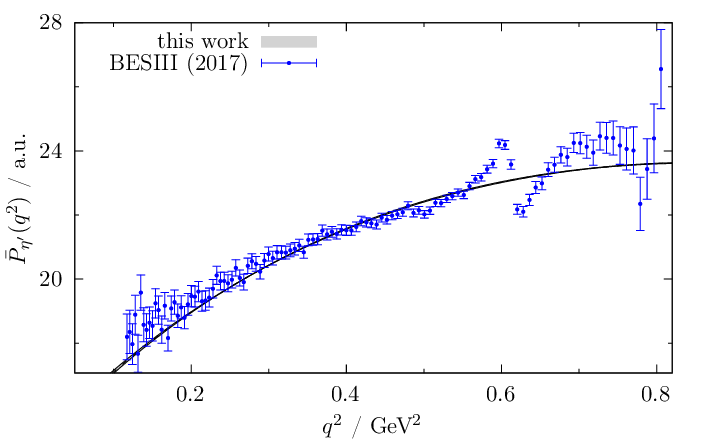}
     \caption{Fits to the $\pi^+\pi^-$ spectra of $\etapp \to \pi^+ \pi^- \gamma$ from KLOE in case of the $\eta$ (top)~\cite{KLOE:2012rfx} and BESIII for the $\eta'$ (bottom)~\cite{BESIII:2017kyd}. The effects of phase-space functions and the $\rho(770)$ resonance have been removed from the spectra shown here. Moreover, for the $\eta'$ data points in the region of the $\omega$ mass have been excluded, since the isospin-breaking effect of $\rho$--$\omega$ mixing was not incorporated in the representation~\eqref{Eq:etapipig}, see Refs.~\cite{Hanhart:2016pcd,Holz:2022hwz}.}
     \label{fig:etapipig_fit}
 \end{figure}

In order to describe the effects of final-state interactions, we introduce pairwise rescattering of the final-state pions via the Omn\`es function. In particular, for the $a_2$ contribution these final-state interactions induce terms that no longer factorize into products of functions of the photon virtualities, but still fulfill the constraints from unitarity and analyticity. On the technical level, the unitarity relation leads to an inhomogeneous Omn\`es problem, whose solution requires the evaluation along carefully chosen complex paths of the kinematic variables to ensure numerical stability~\cite{Gasser:2018qtg}, see Refs.~\cite{Holz:2022smu,Holz:2024diw} for details. It is for this reason that we utilize a representation of the $\pi\pi$ $P$-wave phase shift from unitarized ChPT~\cite{Dobado:1989qm,Truong:1991gv,Dobado:1992ha,Niehus:2020gmf}, as this can be easily evaluated for complex arguments. Considering the constraints from ChPT, our final representation involves only two free parameters, one governing the momentum dependence and influencing the overall normalization and another one associated with the coupling strength of the left-hand-cut contribution.
These parameters are determined  by fitting Eq.~\eqref{Eq:etapipig} to the spectra of the real photon decays $\etapp \to \pi^+ \pi^- \gamma$, as measured by KLOE~\cite{KLOE:2012rfx} and BESIII~\cite{BESIII:2017kyd}, see Fig.~\ref{fig:etapipig_fit}.  
 
 Next, starting from Eq.~\eqref{Eq:etapipig}, in principle a second dispersion relation in $t$, again based on the dominant $2\pi$ intermediate states, determines the isovector part of the TFF. 
 To preserve the correct asymptotic behavior and facilitate the analytic continuation to the space-like region, it  is advantageous to
 recast this in the form of a double-spectral representation, which, upon symmetrization in the photon virtualities, reads
 \begin{align}
     F_{\etapp}^{(I=1)}(-Q_1^2,-Q_2^2) &= \frac{1}{\pi^2} \int_{4M_\pi^2}^{\Lambda^2}\diff x\, \diff y\,  \frac{ \rho_{\etapp}(x,y)}{(x+Q_1^2)(y+Q_2^2)} \notag\\ 
     &\qquad + (Q_1 \leftrightarrow Q_2), \label{Eq:TFF_I1}
     \end{align}
     with double-spectral density
   \beq  
     \rho_{\etapp}(x,y) = \frac{x \sigma_\pi^3(x)}{192 \pi} \Im \Big\{\big[F_\pi^V(x)\big]^*  \mathcal{F}_{\etapp \pi\pi\gamma}(x,y)\Big\}. 
 \eeq
 The low-energy region of the isoscalar channel, on the other hand, is dominated by the narrow $\omega$ and $\phi$ resonances. The small contribution thereof can be parameterized by a vector-meson-dominance ansatz,
 \begin{align}
     &F_{\etapp}^{(I=0)}(-Q_1^2,-Q_2^2) = \hspace{-0.215cm}\sum\limits_{V\in \lbrace \omega,\phi\rbrace}\hspace{-0.1cm} \frac{w_{\etapp V \gamma} F_{\etapp \gamma \gamma}  M_V^4}{(M_V^2+Q_1^2)(M_V^2+Q_2^2)}, \label{Eq:TFF_I0}
 \end{align}
 with $M_V$ as mass parameters and weight factors $w_{\etapp V \gamma}$~\cite{Hanhart:2013vba,Gan:2020aco} determined from the respective decays.

\begin{figure*}[tb]
     \centering
     \begin{minipage}[c]{0.495\textwidth}
        \centering
        \includegraphics[width=\linewidth]{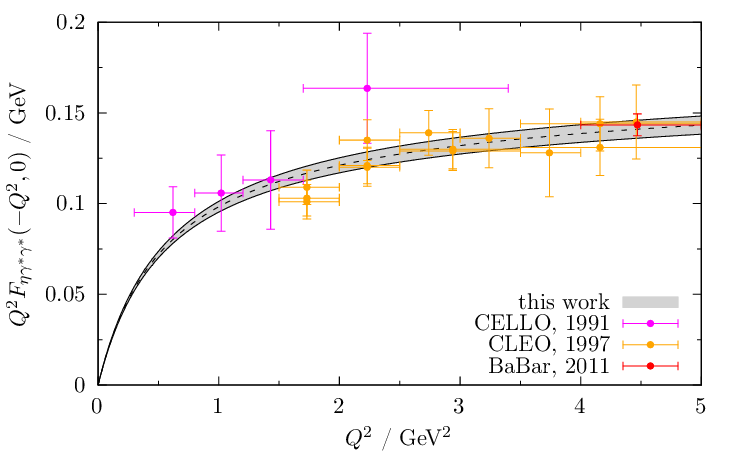}
     \end{minipage}
    \begin{minipage}[c]{0.495\textwidth}
        \centering
        \includegraphics[width=\linewidth]{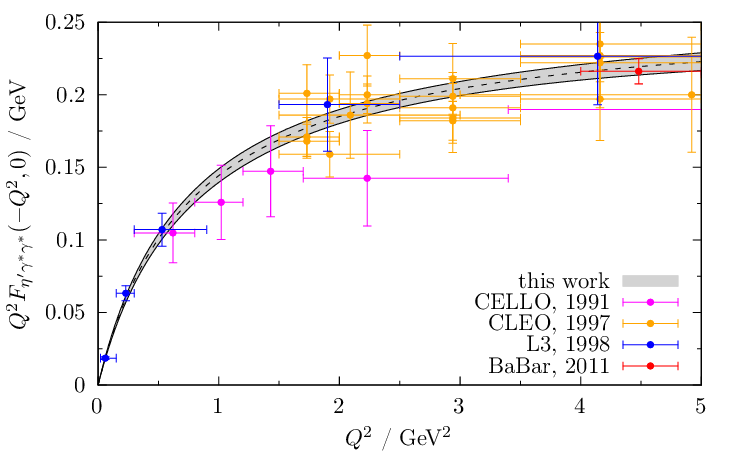}
     \end{minipage}\\
    \begin{minipage}[c]{0.495\textwidth}
        \centering
        \includegraphics[width=\linewidth]{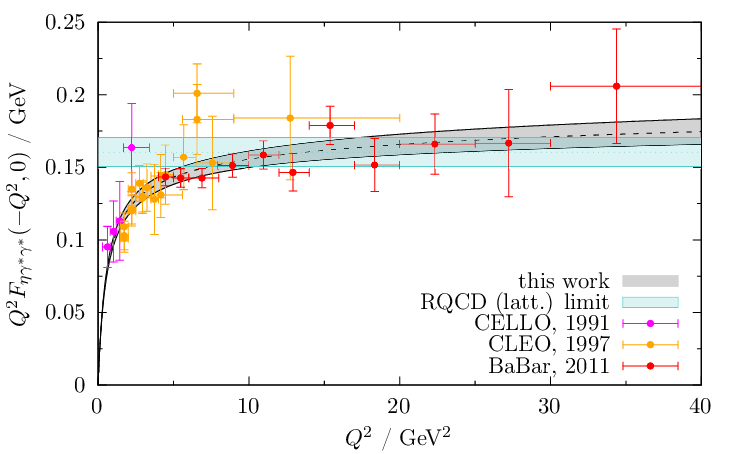}
     \end{minipage}
    \begin{minipage}[c]{0.495\textwidth}
        \centering
        \includegraphics[width=\linewidth]{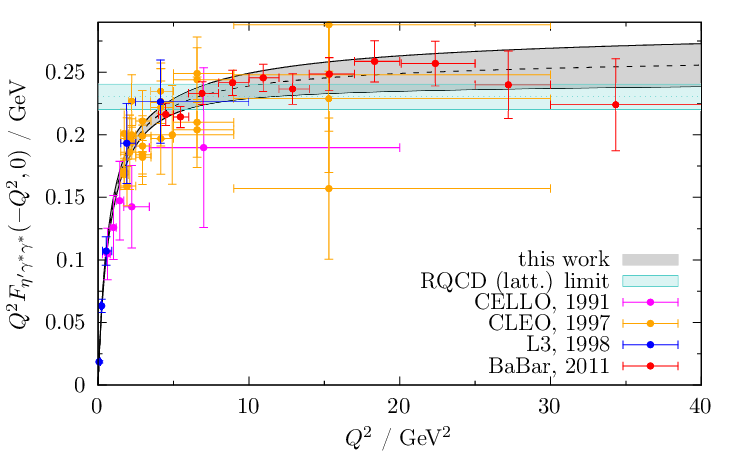}
     \end{minipage}
     \caption{Singly-virtual $\eta$ (left) and $\eta'$ (right) TFFs in the space-like region in comparison to data~\cite{CELLO:1990klc,CLEO:1997fho,L3:1997ocz,BaBar:2011nrp} and the asymptotic limits as extracted from lattice QCD~\cite{Bali:2021qem}. Note that only data points with $Q^2\geq 5\GeV^2$ are included in the fit, while the low-energy region in the upper panel is a prediction.}
     \label{fig:singly_virtual}
 \end{figure*}

 The effective-pole contribution in Eq.~\eqref{Eq:tff_compl} subsumes the effect of higher
 intermediate states, with parameters determined by the sum rule for the normalizations $F_{\etapp\gamma\gamma}$ and a fit to high-energy, space-like data for $e^+e^-\to e^+e^-\etapp$. In practice, we employ the two parameterizations
 \begin{align}
     F_{\etapp}^{\text{eff}\,(A)}(-Q_1^2,-Q_2^2) &= \frac{g_\text{eff} F_{\etapp \gamma \gamma}   M_{\text{eff}}^4}{(M_{\text{eff}}^2+Q_1^2)(M_{\text{eff}}^2+Q_2^2)},\ \label{Eq:TFF_eff}\\
     F_{\etapp}^{\text{eff}\,(B)}(-Q_1^2,-Q_2^2) &= \sum\limits_{V\in \lbrace \rho',\rho''\rbrace}\hspace{-0.1cm} \frac{g_V F_{\etapp \gamma \gamma} M_V^4}{(M_V^2+Q_1^2)(M_V^2+Q_2^2)},\notag
 \end{align}
 where in variant $(A)$ the residue $g_\text{eff}$ is used to restore the normalizations, while $M_\text{eff}$ is fit to singly-virtual TFF data for $Q^2 \geq 5 \GeV^2$. 
 For the $\eta$, the combination of isovector and isoscalar pieces already overfulfills the sum rule, leading to $g_\text{eff}$ in the range $-2\,\%$ to $-13\,\%$, while $M_\text{eff}$ comes out between $(1.3\text{--}2.2)\GeV$. For the $\eta'$, the residue and mass parameters are found around $5\,\%$ and $1.4\,\GeV$, respectively. For both mesons, these parameter ranges support the interpretation of the effective pole as a resummation of higher intermediate states. 
 Variant $(B)$ describes a two-pole structure, where the mass parameters of the $\rho(1450)\equiv\rho'$ and $\rho(1700)\equiv\rho''$ are taken from the Review of Particle Physics (RPP)~\cite{ParticleDataGroup:2024cfk}. One of their residues is employed to fulfill the normalization sum rule, the other one being used as a fit parameter. The fit quality for both variants is quite similar, with reduced $\chi^2 \simeq 1$, see Fig.~\ref{fig:singly_virtual} (lower panel) for the comparison to the input data, $Q^2\geq 5\GeV^2$.  

 In the limit of vanishing mass, the asymptotic piece in Eq.~\eqref{Eq:tff_compl} takes the same form as in Refs.~\cite{Hoferichter:2018dmo,Hoferichter:2018kwz}
 \beq
     F^{\text{asym}}_{\etapp}(q_1^2,q_2^2) = \bar{F}^{\etapp}_{\text{asym}} \int_{\sm}^{\infty} \diff x\, \frac{q_1^2 q_2^2}{(x-q_1^2)^2(x-q_2^2)^2}, \label{Eq:TFF_asym}
 \eeq
which originates from a dispersive reformulation~\cite{Khodjamirian:1997tk} of the leading term of the light-cone expansion evaluated with asymptotic distribution amplitudes~\cite{Lepage:1979zb,Lepage:1980fj,Brodsky:1981rp,Novikov:1983jt,Nesterenko:1982dn,Gorsky:1987idk}. Instead of Eq.~\eqref{Eq:TFF_asym}, we use a variant that includes effects of the pseudoscalar mass~\cite{Zanke:2021wiq,Holz:2024diw}. The asymptotic constraints on the TFFs in the singly- and doubly-virtual directions
 \begin{align}
     \lim\limits_{Q^2\to\infty} Q^2 F_{\etapp \gamma^* \gamma^*}(-Q^2,0) &= \bar{F}^{\etapp}_{\text{asym}},\notag\\
     \lim\limits_{Q^2\to\infty} Q^2 F_{\etapp \gamma^* \gamma^*}(-Q^2,-Q^2) &= \frac{1}{3}\bar{F}^{\etapp}_{\text{asym}},
 \end{align}
 are then fulfilled by our representation~\eqref{Eq:tff_compl}, where ab-initio predictions of the asymptotic value $\bar{F}^{\etapp}_{\text{asym}}$ would depend on the $\eta$--$\eta'$ mixing pattern~\cite{Escribano:2015yup,Gan:2020aco}. However, the asymptotic form~\eqref{Eq:TFF_asym} is constructed in such a way that it does not contribute to the singly-virtual limit, whose coefficient is instead determined by the low-energy pieces~\eqref{Eq:TFF_I1}, \eqref{Eq:TFF_I0} and the effective-pole term~\eqref{Eq:TFF_eff}, and can thus be extracted from the data.
  As an additional source of uncertainty, we also consider input for the asymptotic coefficients from a recent lattice-QCD calculation~\cite{Bali:2021qem}, see lower panel of Fig.~\ref{fig:singly_virtual}.
 
 In order to assess the uncertainties, we proceed as follows:
 (i) ``norm'': the normalizations $F_{\etapp\gamma\gamma}$ are taken from the RPP~\cite{ParticleDataGroup:2024cfk}; 
 (ii) ``disp'': we scan over 
different variants of the dispersive $(I=1)$ piece, involving different integral cutoffs in the range $\Lambda=(1.5\text{--}2.5)\GeV$ and modifications of the subtraction polynomial related to the four-pion-decay amplitude;
(iii) ``BL'': the sensitivity to the singly-virtual Brodsky--Lepage (BL) limit is represented by
the fit parameter in the effective-pole term, which is varied within its fit error ($\simeq 10\,\%$), in addition to the difference between the two variants in Eq.~\eqref{Eq:TFF_eff};
(iv) ``asym'': we vary the threshold parameter in Eq.~\eqref{Eq:TFF_asym} in the range $\sm=1.5(3)\GeV^2$ ($\eta'$) and $\sm=1.4(4)\GeV^2$ ($\eta$), motivated by light-cone-sum-rule calculations~\cite{Khodjamirian:1997tk,Agaev:2014wna} and a smooth matching in the doubly-virtual direction, and consider the variation between data-driven and lattice-QCD determinations of $\bar{F}^{\etapp}_{\text{asym}}$.

As a first application that probes the low-energy properties of the TFF, we study the slope parameter
 \beq
    b_{\etapp} \equiv \frac{1}{F_{\etapp\gamma\gamma}} \frac{\partial}{\partial q^2}  F_{\etapp \gamma^* \gamma^*}(q^2,0) \big|_{q^2=0},
 \eeq
 for which we obtain
 \begin{align}
     b_{\eta} &= 1.833\,(16)_\text{norm}\,(36)_\text{disp}\,(9)_\text{BL}\, [41]_\text{tot}\GeV^{-2},\notag\\
     b_{\eta'} &= 1.493\,(10)_\text{norm}\,(30)_\text{disp}\,(6)_\text{BL}\,[32]_\text{tot}\GeV^{-2},
 \end{align}
 with the total uncertainty referring to the quadratic sum. For both $\eta$ and $\eta'$, the outcome is largely consistent with experimental results based on space-like~\cite{TPCTwoGamma:1990dho,CELLO:1990klc,CLEO:1997fho,L3:1997ocz} and time-like~\cite{Dzhelyadin:1979za,Dzhelyadin:1980kh,NA60:2009una,Usai:2011zza,Berghauser:2011zz,BESIII:2015zpz,Adlarson:2016hpp,BESIII:2024awu,BESIII:2024pxo} data, as well as previous theoretical analyses~\cite{Bramon:1981sw,Ametller:1983ec,Pich:1983zk,Brodsky:1981rp,Ametller:1991jv,Escribano:2013kba,Escribano:2015nra,Escribano:2015yup,Hanhart:2013vba,Holz:2022hwz}. In particular, our result for $b_{\eta'}$ comes out close to $b_{\eta'} = 1.431(23)\GeV^{-2}$~\cite{Holz:2022hwz}, using a highly optimized representation for the singly-virtual TFF including a full treatment of the isospin-breaking $\rho$--$\omega$ mixing correction.  
 
 \emph{Results for $a_\mu$}---Based on these results for the $\etapp$ TFFs, we can now evaluate the corresponding pole contributions to $a_\mu$
 \begin{align}
   a_\mu^{\eta\text{-pole}}&=14.72(56)_\text{norm}(32)_\text{disp}(23)_\text{BL}(54)_\text{asym}[87]_\text{tot},\notag\\
   a_\mu^{\eta'\text{-pole}}&=13.50(48)_\text{norm}(15)_\text{disp}(20)_\text{BL} (48)_\text{asym}[72]_\text{tot},
   \label{Eq:amu}
 \end{align}
 both given in units of $10^{-11}$. The uncertainties are propagated from the TFFs for each of the different error components. Within uncertainties, these results are in agreement with previous analyses, see, e.g., Refs.~\cite{Masjuan:2017tvw,ExtendedTwistedMass:2022ofm,Gerardin:2023naa,Czyz:2017veo,Guevara:2018rhj,Estrada:2024cfy,Eichmann:2019tjk,Raya:2019dnh,Hong:2009zw,Leutgeb:2019zpq}, but significantly more precise.
 A substantial part of the remaining uncertainty in Eq.~\eqref{Eq:amu} derives from the normalizations, e.g., $F_{\eta\gamma\gamma}$ is obtained as an average of extractions from $e^+e^-\to e^+e^-\eta$~\cite{JADE:1985biu,CrystalBall:1988xvy,Roe:1989qy,Baru:1990pc,KLOE-2:2012lws}. This determination could be cross-checked in the future in the JLab Primakoff program~\cite{Gan:2014pna}, especially in view of an inconclusive previous measurement~\cite{Browman:1974sj,Rodrigues:2008zza}. For $F_{\eta'\gamma\gamma}$, we use the fit result from the RPP~\cite{ParticleDataGroup:2024cfk}, which is consistent with the average of $e^+e^-\to e^+e^-\eta'$~\cite{CrystalBall:1988xvy,TPCTwoGamma:1988izb,Roe:1989qy,Butler:1990vv,Baru:1990pc,CELLO:1990klc,CrystalBall:1991zkb,L3:1997ocz}, but includes complementary constraints from other $\eta'$ decays. The asymptotic uncertainty quantifies the sensitivity to the high-energy, doubly-virtual region that is not yet probed sufficiently precisely in experiment~\cite{BaBar:2018zpn}. Here, more detailed comparisons to lattice-QCD calculations could further corroborate or even improve the uncertainties. Likewise, the dispersive error could be consolidated with additional data for $\etapp\to\pi^+\pi^-\gamma$, double-differential data for $e^+e^-\to\etapp\pi^+\pi^-$~\cite{BaBar:2007qju,BaBar:2018erh}, and low-energy, singly-virtual TFF measurements~\cite{BESIII:2020nme}.

 In conclusion, we presented a comprehensive dispersive analysis of the $\etapp$ TFFs, leading to a precision evaluation of the $\etapp$-pole contributions to HLbL scattering in $a_\mu$ with systematically improvable uncertainties. In particular, we included, for the first time, the effects of factorization-breaking contributions generated by the leading $a_2$ left-hand cut, whose dispersive implementation requires the solution of a technically demanding inhomogeneous Omn\`es problem via complex path deformations. Combined with our previous result for the $\pi^0$ pole, we obtain for the sum of the three light pseudoscalars
 \beq
 a_\mu^{\text{PS-poles}}=91.2^{+2.9}_{-2.4}\times 10^{-11},
 \label{Eq:amu_final}
 \eeq
 where the overall uncertainty is now dominated, in fact, by the tension between Belle~\cite{Belle:2012wwz} and BaBar~\cite{BaBar:2009rrj} measurements of the singly-virtual $\pi^0$ TFF at large virtualities, to be scrutinized in the future at Belle II~\cite{Belle-II:2018jsg}. However, even now the uncertainty given in Eq.~\eqref{Eq:amu_final} lies safely below the precision goal set by the final result of the Fermilab experiment.

 \begin{acknowledgments}
 \emph{Acknowledgments}---We thank Judith Plenter for collaboration at early stages of this project.
Financial support by the SNSF (Project Nos.\ 200020\_200553, PCEFP2\_181117, and TMCG-2\_213690) and the DFG through the funds provided to the Sino--German Collaborative
Research Center TRR110 ``Symmetries and the Emergence of Structure in QCD''
(DFG Project-ID 196253076 -- TRR 110) is gratefully acknowledged. 
\end{acknowledgments}

\bibliography{amu}

\end{document}